\documentclass[a4 paper,12pt]{article}
\usepackage{amsmath}
\usepackage{amsfonts}
\usepackage{amssymb}
\usepackage{graphicx}
\usepackage{array}
\usepackage{multirow}
\usepackage{setspace}
\usepackage[margin=1 in]{geometry}
\usepackage{float}
\usepackage[utf8]{inputenc}
\usepackage{xcolor}
\newcolumntype{P}[1]{>{\centering\arraybackslash}p{#1}}
\newcolumntype{M}[1]{>{\centering\arraybackslash}m{#1}}
\usepackage{cite}
\usepackage{hyperref}
\usepackage{authblk}

\begin{document}
\begin{center}
{\bf\large{On the Quantization of FLPR Model} }

\vskip 1.5 cm

{\sf{ \bf Ansha S Nair and Saurabh Gupta}}\\
\vskip .1cm
{\it Department of Physics, National Institute of Technology Calicut,\\ Kozhikode - 673 601, Kerala, India}\\
\vskip .15cm
{E-mails: {\tt anshsuk8@gmail.com, saurabh@nitc.ac.in}}
\end{center}
\vskip 1cm

\noindent
\textbf{Abstract:} We quantize the Friedberg-Lee-Pang-Ren (FLPR) model, using an admissible gauge condition, within the framework of modified Faddeev-Jackiw formalism. Further, we deduce the gauge symmetries and establish off-shell nilpotent and absolutely anti-commuting (anti-)BRST symmetries. 
We also show that the physical states of the theory are annihilated by the first class constraints which is consistent  \textit{\`{a} la} Dirac formalism. 

\clearpage

\section{Introduction}
Systems endowed with singular Lagrangian are called constrained Hamiltonian systems or degenerate systems. Gauge field theories, super-symmetric theories and gravitational field theories are some of the known examples of constrained theories. Dirac and Bergmann pioneered the theoretical studies in constrained systems and its quantization \cite{b,1,2,3,4}.  
Alternatively, Faddeev and Jackiw proposed a formalism which is geometrically motivated and based on the symplectic structure of the phase space to deal with constraints \cite{5}. This formalism is applicable to the first-order Lagrangians and the constraints are deduced with the aid of the zero-modes of the symplectic two-form matrix. To derive new constraints in the system a consistency condition on constraints together with symplectic equations of motion is being used -- which is named as modified Faddeev-Jackiw formalism \cite{9,11}. These constraints are then introduced into the Lagrangian by means of Lagrange multipliers, in an iterative manner, until all the constraints are deduced. The basic brackets are obtained from the components of inverse of the endmost symplectic two-form matrix \cite{6,7,8,SG1,SG2,10}.

Unlike the Dirac method, the Faddeev-Jackiw formalism does not introduce any velocity independent expression as constraints in the theory \cite{Jack}.  This method is more intuitive as it fixes the gauge implicitly \cite{Ram}. Moreover, the obtained symplectic potential, at the final stage of the iteration, is exactly same as the Hamiltonian of the system which is derived through a tedious procedure in the Dirac method (cf. \cite{7} for details).
Recently, Faddeev-Jackiw formalism has been employed to study diverse systems such as massless quantum electrodynamics in the eikonal limit \cite{FJ1}, a bigravity model in one dimension \cite{FJ2}, noncommutative Podolsky theory \cite{FJ3}, Bonzom–Livine model describing gravity in three
dimensions \cite{FJ4}, three-dimensional topologically massive ADS gravity \cite{FJ5}, and also in general relativity and extended gravity \cite{FJ6}.

\par 
However, few interesting and important models are still there to be investigated within the framework of modified Faddeev-Jackiw formalism. 
A soluble gauge model of single non-relativistic particle of unit mass with the characteristics of Gribov ambiguity is one such model. This model is introduced by Friedberg, Lee, Pang and Ren, thus, known as FLPR model \cite{12}. This model is being studied using Hamiltonian and path integral formulation and an alternative method is proposed which allows the inclusion of all gauge copies \cite{12} as contrary to Gribov's suggestion \cite{13}. Further, a detailed Becchi-Rouet-Stora-Tyutin (BRST) analysis, by summing over all Gribov-type copies, was carried out in \cite{14}. The FLPR  model is also being analysed using gauge independent method of abstracting the reduced physical space and complications related to gauge fixing are studied \cite{15}. An investigation on the advantages of using a physical projector, in the quantisation of gauge invariant system, is as well carried out in the context of FLPR model \cite{16}.

\par Thus, our main motive of the present work is to quantize the FLPR model and deduce the constraint structure, by choosing a well defined space-axial gauge condition which does not suffer any Gribov-type copies, through a geometrically motivated approach. Second, we wish to obtain all the basic brackets and the gauge transformations in the theory. Finally, we plan to establish the (anti-)BRST symmetries of the theory.

\par Our present manuscript is organized as follows. Section \ref{2} deals with a detailed analysis of the constraint structure and quantization of the FLPR model, in Cartesian coordinates, within the framework of modified Faddeev-Jackiw formalism. In addition, we establish the gauge symmetries and off-shell nilpotent and absolutely anti-commuting (anti-)BRST symmetries of the system. Similar analysis of the FLPR model, in polar coordinates, is carried out in section \ref{3}. Finally, in section \ref{4}, we summarise our results.
In Appendix 1, we provide the explicit calculations of Dirac brackets and constraint structure via Dirac formalism. Whereas, Appendix 2 comprises of the physical interpretation for the Lagrange multipliers that appear in Faddeev-Jackiw formalism.

\section{FLPR Model in Cartesian Coordinates}\label{2}
The Lagrangian of FLPR model in Cartesian coordinates is given by \cite{12}
\begin{equation}\label{Lc}
    L=\frac{1}{2}\Big[(\dot{x}+gy\zeta)^{2}+(\dot{y}-gx\zeta)^{2}+(\dot{z}-\zeta)^{2}\Big]-U(x^{2}+y^{2}),
\end{equation}
where $\dot{x}$, $\dot{y}$ and $\dot{z}$ denote the generalised velocities corresponding to the generalised coordinates $x$, $y$ and $z$ respectively. Here, $\zeta$ is a gauge variable and $g>0$ represents the coupling constant. The canonical conjugate momenta corresponding to these variables are obtained as follows:
\begin{equation}\label{Cm}
    P_{x}=\dot{x}+gy\zeta, \qquad P_{y}=\dot{y}-gx\zeta, \qquad P_{z}=\dot{z}-\zeta, \qquad P_{\zeta}=0.
\end{equation}
The canonical Hamiltonian $(H_{c})$ of the system is given as
\begin{equation}\label{Hc}
    H_{c}=\frac{1}{2}\Big(P_{x}^{2}+P_{y}^{2}+P_{z}^{2}\Big)+\zeta\Big(g(xP_{y}-yP_{x})+P_{z}\Big)+U(x^{2}+y^{2}),
\end{equation}
which can be derived from Lagrangian \eqref{Lc} using Legendre transformations.

\subsection{Faddeev-Jackiw Quantization}\label{1.1}
To make use of the Faddeev-Jackiw formalism, we express the Lagrangian (\ref{Lc}) in the first-order form using momenta as auxiliary variables in the following fashion
\begin{equation}\label{4eq}
    L_{f}^{(0)}=P_{x}\dot{x}+P_{y}\dot{y}+P_{z}\dot{z}-V^{(0)},
\end{equation}
where
\begin{equation}
    V^{(0)}=\frac{1}{2}\Big(P_{x}^{2}+P_{y}^{2}+P_{z}^{2}\Big)+\zeta\Big(g(xP_{y}-yP_{x})+P_{z}\Big)+U(x^{2}+y^{2}).
\end{equation}
The corresponding symplectic equations of motion can be written as
\begin{equation}\label{6eq}
    f_{ij}^{(0)}{\dot{\chi}^{j}}=\frac{\partial V^{(0)}(\chi)}{\partial\chi^{i}},
\end{equation}
where the antisymmetric symplectic matrix $f_{ij}^{(0)}$ is defined as
\begin{equation}\label{7}
    f_{ij}^{(0)}=\frac{\partial a_{j}(\chi)}{\partial\chi^{i}}-\frac{\partial a_{i}(\chi)}{\partial\chi^{j}},\qquad \text{with} \quad  a_{i}=\frac{\partial L}{\partial{\dot{\chi}^{i}}}.
\end{equation}
We choose the set of symplectic variables as
\begin{equation}
    \chi^{(0)}=\{x, P_{x}, y, P_{y}, z, P_{z}, \zeta\},
\end{equation}
and with this choice, the components of symplectic one-form turn out to be
\begin{equation}
    a_{x}^{(0)}=P_{x},\quad a_{P_{x}}^{(0)}=0,\quad a_{y}^{(0)}=P_{y},\quad a_{P_{y}}^{(0)}=0,\quad
    a_{z}^{(0)}=P_{z},\quad a_{P_{z}}^{(0)}=0,\quad a_{\zeta}^{(0)}=0.
\end{equation}
Thus, the symplectic matrix $(f_{ij}^{(0)})$, using \eqref{7}, takes the form  
\begin{equation}
f_{ij}^{(0)}=\begin{pmatrix}
0&-1&0&0&0&0&0\\
1&0&0&0&0&0&0\\
0&0&0&-1&0&0&0\\
0&0&1&0&0&0&0\\
0&0&0&0&0&-1&0\\
0&0&0&0&1&0&0\\
0&0&0&0&0&0&0
\end{pmatrix},
\end{equation}
which is apparently singular. The zero-mode $(\nu^{(0)}\big)^{T}$of this two-form symplectic matrix $(f_{ij}^{(0)})$ is
\begin{equation}
    \big(\nu^{(0)}\big)^{T}=\begin{pmatrix}
        0&0&0&0&0&0&\nu_{\lambda}
    \end{pmatrix},
\end{equation}
where $\nu_{\lambda}$ is an arbitrary constant. The constraints in the theory can be obtained by contracting this zero-mode with the equations of motion given in \eqref{6eq}, in the following fashion
\begin{equation}\label{12}
    \Omega^{(0)}\equiv(\nu^{(0)})^{T}\frac{\partial V^{(0)}(\chi)}{\partial \chi^{(0)}}=0,
\end{equation}
where,
\begin{equation}
    \frac{\partial V^{(0)}(\chi)}{\partial \chi^{(0)}}=\begin{pmatrix}
        \zeta gP_{y}+2U^{'}x\\P_{x}-\zeta gy\\-\zeta gP_{x}+2U^{'}y\\P_{y}+\zeta gx\\0\\P_{z}+\zeta\\g(xP_{y}-yP_{x})+P_{z}
    \end{pmatrix}.
\end{equation}
Here $U^{'}$ represents the derivative of $U$ with respect to $(x^2 + y^2)$.
Thus, the explicit form of the constraint is given as
\begin{equation}\label{cc}
    \Omega^{(0)}\equiv\nu_{\lambda}\Big(g(xP_{y}-yP_{x})+P_{z}\Big).
\end{equation}
Further, in order to deduce new constraints in the system, we make use of the modified Faddeev-Jackiw formalism \cite{9,SG1}. Consequently, applying consistency condition on constraints analogous to the Dirac-Bergmann algorithm as follows
\begin{equation}\label{15}
    {\dot{\Omega}^{(0)}}=\frac{\partial\Omega^{(0)}}{\partial\chi^{i}}{\dot{\chi}^{i}}=0.
\end{equation}
Combining \eqref{6eq} and \eqref{15}, we get
\begin{equation}\label{16}
    \bar{f}_{kj}^{(1)}{\dot{\chi}}^{j}=Z_{k}(\chi),
\end{equation}
where
\begin{equation}
     \bar{f}_{kj}^{(1)}=\begin{pmatrix}
         f_{ij}^{(0)}\\
         \frac{\partial\Omega^{(0)}}{\partial\chi^{i}}
     \end{pmatrix},\qquad
     Z_{k}(\chi)=\begin{pmatrix}
             \frac{\partial V^{(0)}(\chi)}{\partial\chi^{i}} \\0
         \end{pmatrix}.
             \end{equation}
The symplectic matrix $\bar{f}_{kj}^{(1)}$ is evaluated as follows
\begin{equation}
    \bar{f}_{kj}^{(1)}=\begin{pmatrix}
        0&-1&0&0&0&0&0\\
1&0&0&0&0&0&0\\
0&0&0&-1&0&0&0\\
0&0&1&0&0&0&0\\
0&0&0&0&0&-1&0\\
0&0&0&0&1&0&0\\
0&0&0&0&0&0&0\\
gP_{y}&-gy&-gP_{x}&gx&0&1&0
    \end{pmatrix},
\end{equation}
which is not a square matrix. However, the zero-mode $(\bar{\nu}^{(1)})^{T}$ of this matrix is
\begin{equation}
    \Big(\bar{\nu}^{(1)}\Big)^{T}=
    \begin{pmatrix}
        -gy\quad-gP_{y}\quad gx\quad gP_{x}\quad 1\quad0\quad\nu_{\lambda}^1\quad 1\end{pmatrix},
\end{equation}
where $\nu_{\lambda}^1$ is an arbitrary constant. Multiplying this zero-mode to the both sides of \eqref{16} and evaluating with the condition $\Omega^{(0)}=0$ give rise to new constraints in the theory. The condition $\Omega^{(0)}=0$ is to ensure that the obtained 
 constraints do not appear in the following calculation \cite{10}. So, we have
\begin{equation}
    (\bar{\nu}^{(1)})^{T}Z_{k}\vline_{\Omega^{(0)}=0}=0,
\end{equation}
which turns out to be an identity. Therefore, we infer that there are no further constraints in the theory. Now, we introduce the obtained constraint into the Lagrangian using a Lagrange multiplier $(\beta)$ as
\begin{equation}
    L_{f}^{(1)}=P_{x}\dot{x}+P_{y}\dot{y}+P_{z}\dot{z}+(g(xP_{y}-yP_{x})+P_{z})\dot{\beta}-V^{(1)},
\end{equation}
where
\begin{equation}
    V^{(1)}=V^{(0)}\vline_{\Omega^{(0)}=0}=\frac{1}{2}\Big(P_{x}^{2}+P_{y}^{2}+P_{z}^{2}\Big)+U(x^{2}+y^{2}).
\end{equation}
We treat $\beta$ as a symplectic variable and hence the set of first-iterated symplectic variables  $(\chi^{(1)})$ becomes
\begin{equation}
    \chi^{(1)}=\{x, P_{x}, y, P_{y}, z, P_{z}, \beta\}.
\end{equation}
The corresponding components of symplectic one-forms can be evaluated as
\begin{eqnarray}
    a_{x}^{(1)}=P_{x},\qquad a_{P_{x}}^{(1)}=0,\qquad a_{y}^{(1)}=P_{y},\qquad a_{P_{y}}^{(1)}=0, \nonumber \\
    a_{z}^{(1)}=P_{z},\qquad a_{P_{z}}^{(1)}=0,\qquad a_{\beta}^{(1)}=g(xP_{y}-yP_{x})+P_{z}.
\end{eqnarray}
Consequently, the first-iterated symplectic matrix 
\begin{equation}
    f_{ij}^{(1)}=\begin{pmatrix}
        0&-1&0&0&0&0&gp_{y}\\
1&0&0&0&0&0&-gy\\
0&0&0&-1&0&0&-gP_{x}\\
0&0&1&0&0&0&gx\\
0&0&0&0&0&-1&0\\
0&0&0&0&1&0&1\\
-gP_{y}&gy&gP_{x}&-gx&0&-1&0
    \end{pmatrix},
\end{equation}
 turns out to be a singular matrix. This implies that the underlying theory is a gauge theory. Further, in order to quantize the system, we choose an admissible gauge condition $z=0$ \cite{12,14}, and incorporate this gauge condition into the first-iterated Lagrangian by means of a Lagrange multiplier, say $\alpha$, as
\begin{equation}\label{26}
    L_{f}^{(2)}=P_{x}\dot{x}+P_{y}\dot{y}+P_{z}\dot{z}+(g(xP_{y}-yP_{x})+P_{z})\dot{\beta}+z\dot{\alpha}-V^{(2)},
\end{equation}
where
\begin{equation}
    V^{(2)}=V^{(1)}\vline_{z=0}=\frac{1}{2}\Big(P_{x}^{2}+P_{y}^{2}+P_{z}^{2}\Big)+U(x^{2}+y^{2}).
\end{equation}
Now, the set of second-iterated symplectic variables becomes
\begin{equation}
    \chi^{(2)}=\{x, P_{x}, y, P_{y}, z, P_{z}, \beta,\alpha\},
\end{equation}
and the respective components of symplectic one-forms can be given as
\begin{eqnarray}
&&    a_{x}^{(2)}=P_{x},\qquad a_{P_{x}}^{(2)}=0,\qquad a_{y}^{(2)}=P_{y},\qquad a_{P_{y}}^{(2)}=0, \nonumber \\
&&    a_{z}^{(2)}=P_{z},\qquad a_{P_{z}}^{(2)}=0,\qquad a_{\beta}^{(2)}=g(xP_{y}-yP_{x})+P_{z},\qquad a_{\alpha}^{(2)}=z.
\end{eqnarray}
Subsequently, we obtain the second-iterated symplectic matrix as
\begin{equation}
    f_{ij}^{(2)}=\begin{pmatrix}
        0&-1&0&0&0&0&gp_{y}&0\\
1&0&0&0&0&0&-gy&0\\
0&0&0&-1&0&0&-gP_{x}&0\\
0&0&1&0&0&0&gx&0\\
0&0&0&0&0&-1&0&1\\
0&0&0&0&1&0&1&0\\
-gP_{y}&gy&gP_{x}&-gx&0&-1&0&0\\
0&0&0&0&-1&0&0&0
    \end{pmatrix},
\end{equation}
which is a non singular matrix. The inverse of the matrix $(f_{ij}^{(2)})^{-1}$can be evaluated as
\begin{equation}
    \Big(f_{ij}^{(2)}\Big)^{-1}=\begin{pmatrix}
        0&1&0&0&0&gy&0&gy\\
-1&0&0&0&0&gP_{y}&0&gP_{y}\\
0&0&0&1&0&-gx&0&-gx\\
0&0&-1&0&0&-gP_{x}&0&-gP_{x}\\
0&0&0&0&0&0&0&-1\\
-gy&-gP_{y}&gx&gP_{x}&0&0&-1&0\\
0&0&0&0&0&1&0&1\\
-gy&-gP_{y}&gx&gP_{x}&1&0&-1&0
    \end{pmatrix}.
\end{equation}
Thus, the basic brackets in the theory, \textit{\`{a} la} Faddeev-Jackiw formalism, can be directly obtained from the components of the inverse matrix $(f_{ij}^{(2)})^{-1}$ as \cite{5}
\begin{equation}
    \Big{\{}\chi_{i}^{(2)}, \chi_{j}^{(2)}\Big{\}}=\Big(f_{ij}^{(2)}\Big)^{-1}.
\end{equation}
In explicit form, these brackets are given as:
\begin{eqnarray}\label{33}
  &&\big\{x,P_{x}\big\}=1=-\big\{P_{x},x\big\},\quad
     \big\{x,P_{z}\big\}=gy=-\big\{P_{z},x\big\},\quad
      \big\{x,\alpha\big\}=gy=-\big\{\alpha,x\big\}, \nonumber\\
      &&\big\{y,P_{y}\big\}=1=-\big\{P_{y},y\big\},\quad
     \big\{P_{z},y\big\}=gx=-\big\{y,P_{z}\big\},\quad
      \big\{\alpha,y\big\}=gx=-\big\{y,\alpha\big\},\nonumber\\\
      &&\big\{\alpha,z\big\}=1=-\big\{z,\alpha\big\},\quad
            \big\{P_{x},P_{z}\big\}=gP_{y}=-\big\{P_{z},P_{x}\big\},\quad
     \big\{\beta,\alpha\big\}=1=-\big\{\alpha,\beta\big\},\nonumber\\
      &&\big\{P_{z},P_{y}\big\}=gP_{x}=-\big\{P_{y},P_{z}\big\},\qquad
      \big\{\alpha, P_y \big\}=gP_{x}=-\big\{P_{y}, \alpha\big\},\nonumber\\
       && \big\{P_{x},\alpha\big\}=gP_{y}=-\big\{\alpha,P_{x}\big\}, \qquad\big\{\beta,P_{z}\big\}=1=-\big\{P_{z},\beta\big\}.
\end{eqnarray}
All the brackets among the dynamical variables in the theory coincide with the Dirac brackets obtained through Dirac's quantization procedure (cf. Appendix 1 for detailed calculation of Dirac brackets).

\subsection{Gauge and (anti-)BRST Symmetries}
As we have seen that after incorporating all the constraints into the Lagrangian, the symplectic matrix $(f_{ij}^{(1)})$ still remains singular. This indicates that our system has a gauge symmetry and the zero-mode of the corresponding matrix acts as the generator of the gauge transformations \cite{FJ2}. Keeping this in mind, we calculate the zero-mode $(\nu^{(1)})^{T}$of the symplectic matrix  $(f_{ij}^{(1)})$, as
\begin{equation}
    \big(\nu^{(1)}\big)^{T}=\begin{pmatrix}
        gy&gP_{y}&-gx&-gP_{x}&-1&0&1
    \end{pmatrix},
\end{equation}
which acts as the generator of the gauge transformations $(\delta)$ in the following fashion \cite{8}
\begin{equation}
    \delta\chi_{k}^{1}=\nu_{k}^{1}\lambda(t),
    \end{equation}
where $\chi_{k}^{1}$ is the set of symplectic variables and $\lambda(t)$ denotes the time-dependent infinitesimal gauge parameter. 
Explicitly, these transformations are listed as 
\begin{equation}\label{gtc}
\begin{split}
    &\delta x=gy\lambda(t),\qquad \delta P_{x}=gP_{y}\lambda(t),\qquad
    \delta y=-gx\lambda(t),\\ &\delta P_{y}=-gP_{x}\lambda(t),\qquad
    \delta z=-\lambda(t),\qquad \delta P_{z}=0, \qquad \delta\zeta=-{\dot{\lambda}(t)}.
    \end{split}
\end{equation}
It is straightforward to verify that the first-order Lagrangian $L_{f}^{(0)}$ remains invariant under this set of gauge transformations.

Any gauge invariant theory can be rewrittten as a quantum theory that possesses a generalized gauge symmetry called the BRST symmetry \cite{17}.  Accordingly, one enlarges the Hilbert space of gauge invariant theory by introducing anti-commuting variables $(\Bar{\cal{C}})\cal{C}$ called the Faddeev-Popov (anti-)ghost variables, and a commuting variable $b$ called the Nakanishi-Lautrup variable. Further, the form of gauge transformation is replaced by a BRST transformation which mixes the operators having different statistics. To this end, we explore the gauge theory of FLPR model in the framework of BRST formalism \cite{18}. Thus, the (anti-)BRST 
$(\delta_{ab}) \delta_{b}$  symmetry transformations  can be written as
\begin{equation}\label{37}
\begin{split}
    &\delta_{b} x=gy{\cal{C}},\quad \delta_{b} P_{x}=gP_{y}{\cal{C}},\quad
    \delta_{b} y=-gx{\cal{C}},\quad \delta_{b} P_{y}=-gP_{x}{\cal{C}},\quad \delta_{b} z=-{\cal{C}},\\&
     \delta_{b} P_{z}=0, \qquad \delta_{b}\zeta=-\dot{\cal{C}},\qquad
   \delta_{b}{\cal{C}}=0,\qquad \delta_{b}\bar{\cal{C}}=-b,\qquad\delta_{b} b=0,
    \end{split}
\end{equation}
and
\begin{equation}\label{111}
\begin{split}
    &\delta_{ab} x=gy{\cal\bar{C}},\quad \delta_{ab} P_{x}=gP_{y}{\cal\bar{C}},\quad
    \delta_{ab} y=-gx{\cal\bar{C}},\quad\delta_{ab} P_{y}=-gP_{x}{\cal\bar{C}},\quad\delta_{ab} b=0,\\& 
    \delta_{ab} z=-{\cal\bar{C}},\qquad
     \delta_{ab} P_{z}=0, \qquad \delta_{ab}\zeta=-{\dot{\bar{\cal{C}}}},\qquad
   \delta_{ab}{\cal\bar{C}}=0,\qquad \delta_{ab}{\cal{C}}=b.
    \end{split}
\end{equation}
These (anti-)BRST transformations are off-shell nilpotent and absolutely anti-commuting in nature.
The (anti-)BRST invariant Lagrangian can be constructed by adding an (anti-)BRST invariant function, which embodies the guage fixing term and Faddeev-Popov ghost terms, to the first order Lagrangian $L_{f}^{(0)}$ in the following fashion:
\begin{eqnarray}\label{b}
    L_{b}&=&L_{f}^{(0)}+\delta_{b}\Big(\bar{\cal{C}}(\dot{\zeta}-z-\frac{1}{2}b)\Big)\nonumber\\&\equiv& L_{f}^{(0)}+\delta_{ab}\Big(-{\cal{C}}(\dot{\zeta}-z-\frac{1}{2}b)\Big)\nonumber\\
    &\equiv&P_{x}\dot{x}+P_{y}\dot{y}+P_{z}\dot{z}-\frac{1}{2}\Big(P_{x}^{2}+P_{y}^{2}+P_{z}^{2}\Big)-\zeta\Big(g(xP_{y}-yP_{x})+P_{z}\Big)\nonumber\\&-&U(x^{2}+y^{2})-b(\dot{\zeta}-z-\frac{1}{2}b)-{\dot{\bar{\cal{C}}}}{\dot{\cal{C}}}-{\cal{\bar{C}}}{\cal{C}}.
\end{eqnarray}
Here $({\bar{\cal{C}}}){\cal{C}}$ are the fermionic (anti-)ghost variables and $b$ is the Nakanishi-Lautrup variable which linearize the quadratic gauge fixing term $\frac{1}{2}(\dot{\zeta}-z)^{2}$.
The above Lagrangian remains quasi-invariant under the (anti-)BRST transformations as
\begin{equation}
    \delta_{b}L_{b}=\frac{d}{dt}(b\dot{\cal{C}}), \qquad \delta_{ab}L_{b}=\frac{d}{dt}(b\dot{\bar{\cal{C}}}).
\end{equation}
The conserved (anti-)BRST charges, which are the generators of the (anti-)BRST symmetries given in \eqref{37} and \eqref{111}, can be derived using Noether's theorem as
\begin{equation}
    Q_{b}={\cal{C}}(g(yP_{x}-xP_{y})-P_{z})-\dot{\cal{C}}P_{\zeta}, \qquad Q_{ab}=\bar{\cal{C}}(g(yP_{x}-xP_{y})-P_{z})-\dot{\bar{\cal{C}}}P_{\zeta}.
\end{equation}
The conservation of the (anti-)BRST charges can be proved with the aid of the following Euler-Lagrange equations of motion corresponding to $L_{b}$
\begin{equation}
    \begin{split}
        &\dot{P_{x}}+\zeta gP_{y}+2xU^{'}=0, \qquad \dot{x}-P_{x}+\zeta gy=0,\qquad \dot{P_{y}}-\zeta gP_{x}+2yU^{'}=0,\\ &\dot{y}-P_{y}-\zeta gx=0,\qquad \dot{\zeta}-b-z=0,\qquad \dot{b}-g(xP_{y}-yP_{x})-P_{z}=0,\\& \dot{z}-P_{z}-\zeta=0,\qquad \dot{P_{z}}-b=0,\qquad \ddot{\cal{C}}-{\cal{C}}=0,\qquad {\ddot{\bar{\cal{C}}}}-\bar{\cal{C}}=0.
    \end{split}
\end{equation}
Moreover, the dynamically stable subspace of states $ | \psi \rangle $ satisfies the physicality criteria $Q_{(a)b}| \psi \rangle =0$ leads to the conditions $(g(yP_{x}-xP_{y})-P_{z}) | \psi \rangle=0$ and $P_{\zeta}| \psi \rangle=0$. It is noteworthy that $P_{\zeta}\approx0$ and $(g(yP_{x}-xP_{y})-P_{z})\approx0$ are first class constraints in the system. Consequently, the physical states of the theory are annihilated by the first class constraints which is consistent with the Dirac formalism for quantization (cf. Appendix 1 for details).

\section{FLPR Model in Polar Coordinates}\label{3}
We begin with the following Lagrangian describing the dynamics of the FLPR model in polar coordinates \cite{12,16}
\begin{equation}\label{43}
       \Tilde{L}=\frac{1}{2}\dot{r}^{2}+\frac{1}{2}r^{2}(\dot{\theta}-g\zeta)^{2}+\frac{1}{2}(\dot{z}-\zeta)^{2}-V(r),
\end{equation}
where $\dot{r}$, $\dot{\theta}$ and $\dot{z}$ represent the generalised velocities, $\zeta$ is a gauge variable and $g$ denotes a coupling constant.  The canonical momenta corresponding to $r$, $\theta$, $z$ and $\zeta$ are respectively given, as
\begin{equation}\label{cmp}
    P_{r}=\dot{r},\qquad
    P_{\theta}=r^{2}(\dot{\theta}-g\zeta),\qquad
    P_{z}=\dot{z}-\zeta,\qquad P_{\zeta}=0.
\end{equation} 
The canonical Hamiltonian $(\Tilde{H_{c}})$ of the system is given as
\begin{equation}\label{45}
   \Tilde{H}_{c}= \frac{1}{2}P_{r}^{2}+\frac{1}{2r^{2}}P_{\theta}^{2}+\frac{1}{2}P_{z}^{2}+\zeta(gP_\theta+P_{z})+V(r),
\end{equation}
which can be derived from the Lagrangian \eqref{43} with the aid of Legendre transformations.
\subsection{Faddeev-Jackiw Quantization}

To employ Faddeev-Jackiw formalism, we express the Lagrangian $(\Tilde{L})$ in first order form, as
\begin{equation}
\Tilde{L}_{f}^{(0)}=P_{r}\dot{r}+P_{\theta}\dot{\theta}+P_{z}\dot{z}-\Tilde{V}^{(0)},
\end{equation}
where
\begin{equation}
    \Tilde{V}^{(0)}=\frac{1}{2}P_{r}^{2}+\frac{1}{2r^{2}}P_{\theta}^{2}+\frac{1}{2}P_{z}^{2}+\zeta(gP_\theta+P_{z})+V(r).
\end{equation}
In Faddeev-Jackiw formalism, the equations of motion are derived in terms of symplectic matrix $\Tilde{f}_{ij}^{(0)}$ as follows:
\begin{equation}\label{44}
     \Tilde{f}_{ij}^{(0)}{\dot{\Tilde{\chi}}^{j}}=\frac{\partial \Tilde{V}^{(0)}(\Tilde{\chi})}{\partial\Tilde{\chi}^{i}}.
\end{equation}
We take the set of symplectic variables to be
\begin{equation}
    \Tilde{\chi}^{(0)}=\{r, P_{r}, \theta, P_{\theta}, z, P_{z}, \zeta\}.
\end{equation}
Further, we deduce the components of symplectic one-forms and the corresponding symplectic matrix $(\Tilde{f_{ij}}^{(0)})$ respectively, using \eqref{7} as
\begin{equation}
    \Tilde{a}_{r}^{(0)}=P_{r},\quad \Tilde{a}_{P_{r}}^{(0)}=0,\quad \Tilde{a}_{\theta}^{(0)}=P_{\theta},\quad \Tilde{a}_{P_{\theta}}^{(0)}=0,\quad
    \Tilde{a}_{z}^{(0)}=P_{z},\quad \Tilde{a}_{P_{z}}^{(0)}=0,\quad \Tilde{a}_{\zeta}^{(0)}=0,
\end{equation}
and
\begin{equation}
\Tilde{f}_{ij}^{(0)}=\begin{pmatrix}
0&-1&0&0&0&0&0\\
1&0&0&0&0&0&0\\
0&0&0&-1&0&0&0\\
0&0&1&0&0&0&0\\
0&0&0&0&0&-1&0\\
0&0&0&0&1&0&0\\
0&0&0&0&0&0&0
\end{pmatrix}.
\end{equation}
The symplectic matrix $\Tilde{f}_{ij}^{(0)}$ is obviously a singular matrix which indicates the presence of constraints in the theory. The zero-mode of the above symplectic matrix $\Tilde{f}_{ij}^{(0)}$ turns out to be
\begin{equation}
    \big(\Tilde{\nu}^{(0)}\big)^{T}=\begin{pmatrix}
        0&0&0&0&0&0&\nu_{\mu}
    \end{pmatrix},
\end{equation}
where $\nu_{\mu}$ is an arbitrary constant. This zero-mode is being used to obtain the constraints as
\begin{equation}\label{49}
    \Tilde{\Omega}^{(0)}\equiv(\Tilde{\nu}^{(0)})^{T}\frac{\partial \Tilde{V}^{(0)}(\Tilde{\chi})}{\partial \Tilde{\chi}^{(0)}}=0,
\end{equation}
where
\begin{equation}
    \frac{\partial \Tilde{V}^{(0)}(\Tilde{\chi})}{\partial \Tilde{\chi}^{(0)}}=\begin{pmatrix}
        -\frac{P_{\theta}^{2}}{r^{3}}+\frac{\partial V}{\partial r}\\P_{r}\\0\\ \frac{P_{\theta}}{r^{2}}+\zeta g\\0\\P_{z}+\zeta\\gP_{\theta}+P_{z}
    \end{pmatrix}.
\end{equation}
The constraint evaluated using \eqref{49} is
\begin{equation}\label{cp}
    \Tilde{\Omega}^{(0)}\equiv\nu_{\mu}\big(gP_{\theta}+P_{z}\big).
\end{equation}
Now, we implement the modified Faddeev-Jackiw formalism to deduce new constraints in the theory. Applying consistency condition on obtained constraints as
\begin{equation}\label{52}
    {\dot{\Tilde{\Omega}}^{(0)}}=\frac{\partial\Tilde{\Omega}^{(0)}}{\partial\Tilde{\chi}^{i}}{\dot{\Tilde{\chi}}^{i}}=0.
\end{equation}
Combining equations \eqref{44} and \eqref{52}, we get
\begin{equation}\label{53}
    \Tilde{f}_{kj}^{*(1)}\dot{\Tilde{\chi}}^{j}=\Tilde{Z}_{k}(\Tilde{\chi}),
\end{equation}
where
\begin{equation}
     \Tilde{f}_{kj}^{*(1)}=\begin{pmatrix}
         \Tilde{f}_{ij}^{(0)}\\
         \frac{\partial\Tilde{\Omega}^{(0)}}{\partial\Tilde{\chi}^{i}}
     \end{pmatrix},\qquad
     \Tilde{Z}_{k}(\Tilde{\chi})=\begin{pmatrix}
             \frac{\partial \Tilde{V}^{(0)}(\Tilde{\chi})}{\partial\Tilde{\chi}} \\0
         \end{pmatrix}.
             \end{equation}
The explicit form of symplectic matrix $\Tilde{f}_{kj}^{*(1)}$ is given as follows:
\begin{equation}
    \Tilde{f}_{kj}^{*(1)}=\begin{pmatrix}
        0&-1&0&0&0&0&0\\
1&0&0&0&0&0&0\\
0&0&0&-1&0&0&0\\
0&0&1&0&0&0&0\\
0&0&0&0&0&-1&0\\
0&0&0&0&1&0&0\\
0&0&0&0&0&0&0\\
0&0&0&g&0&1&0
    \end{pmatrix},
\end{equation}
which is a non-square matrix. However, this matrix has following zero-mode 
\begin{equation}
    ({\Tilde{\nu}}^{*(1)})^{T}=
    \begin{pmatrix}
    0&0&g&0&1&0&\nu_{\mu}^{1}&1
        \end{pmatrix},
\end{equation}
where $\nu_{\mu}^{1}$ is an arbitrary constant. As we already mentioned that multiplying this zero-mode $({\Tilde{\nu}}^{*(1)})^{T}$ to \eqref{53} leads to the new constraints in the system. Therefore, we get
\begin{equation}
    ({\Tilde{\nu}}^{*(1)})^{T}\Tilde{Z}_{k}\vline_{\Tilde{\Omega}^{(0)}=0}=0 \implies 0=0.
\end{equation}
This identity implies that there are no further constraints in the system. So, following the Faddeev-Jackiw formalism, the obtained constraint is introduced into the Lagrangian by means of a Lagrange multiplier $(\rho)$ as follows:
\begin{equation}
\Tilde{L}_{f}^{(1)}=P_{r}\dot{r}+P_{\theta}\dot{\theta}+P_{z}\dot{z}+(gP_{\theta}+P_{z})\dot{\rho}-\Tilde{V}^{(1)},
\end{equation}
where
\begin{equation}
    \Tilde{V}^{(1)}=\Tilde{V}^{(0)}\vline_{\Tilde{\Omega}^{(0)}=0}=\frac{1}{2}P_{r}^{2}+\frac{1}{2r^{2}}P_{\theta}^{2}+\frac{1}{2}P_{z}^{2}+V(r).
\end{equation}
Now, the set of first-iterated symplectic variables are chosen as
\begin{equation}
    \Tilde{\chi}^{(1)}=\{r, P_{r}, \theta, P_{\theta}, z, P_{z}, \rho\},
\end{equation}
and accordingly components of symplectic one-forms are obtained in following fashion:
\begin{eqnarray}
    &&\Tilde{a}_{r}^{(1)}=P_{r},\qquad \Tilde{a}_{P_{r}}^{(1)}=0,\qquad \Tilde{a}_{\theta}^{(1)}=P_{\theta},\qquad \Tilde{a}_{P_{\theta}}^{(1)}=0, \nonumber \\
    &&\Tilde{a}_{z}^{(1)}=P_{z},\qquad \Tilde{a}_{P_{z}}^{(1)}=0,\qquad \Tilde{a}_{\rho}^{(1)}=gP_{\theta}+P_{z}.
\end{eqnarray}
Further, we evaluate the first-iterated symplectic matrix as follows:
\begin{equation}
    \Tilde{f}_{ij}^{(1)}=\begin{pmatrix}
        0&-1&0&0&0&0&0\\
1&0&0&0&0&0&0\\
0&0&0&-1&0&0&0\\
0&0&1&0&0&0&g\\
0&0&0&0&0&-1&0\\
0&0&0&0&1&0&1\\
0&0&0&-g&0&-1&0
    \end{pmatrix},
\end{equation}
which is a singular matrix. Singularity of symplectic matrix $\Tilde{f}_{ij}^{(1)}$ and absence of new constraints suggests that underlying theory has a gauge symmetry. Now, to fix the gauge and quantize the theory, we choose a permissible gauge condition $z=0$ \cite{12,14} and introduce it into the Lagrangian with the help of a Lagrange multiplier, say $\sigma$, as 
\begin{equation}\label{65}
\Tilde{L}_{f}^{(2)}=P_{r}\dot{r}+P_{\theta}\dot{\theta}+P_{z}\dot{z}+(gP_{\theta}+P_{z})\dot{\rho}+z\dot{\sigma}-\Tilde{V}^{(2)},
\end{equation}
where
\begin{equation}
    \Tilde{V}^{(2)}=\Tilde{V}^{(1)}\vline_{z=0}=\frac{1}{2}P_{r}^{2}+\frac{1}{2r^{2}}P_{\theta}^{2}+\frac{1}{2}P_{z}^{2}+V(r).
\end{equation}
Now, we choose set of second-iterated symplectic variables to be
\begin{equation}
    \Tilde{\chi}^{(2)}=\{r, P_{r}, \theta, P_{\theta}, z, P_{z}, \rho, \sigma\},
\end{equation}
and the corresponding components of one-forms are
\begin{eqnarray}
    &&\Tilde{a}_{r}^{(2)}=P_{r},\qquad \Tilde{a}_{P_{r}}^{(2)}=0,\qquad \Tilde{a}_{\theta}^{(2)}=P_{\theta},\qquad \Tilde{a}_{P_{\theta}}^{(2)}=0, \nonumber \\
   &&\Tilde{a}_{z}^{(2)}=P_{z},\qquad \Tilde{a}_{P_{z}}^{(2)}=0,\qquad \Tilde{a}_{\rho}^{(2)}=gP_{\theta}+P_{z},\qquad \Tilde{a}_{\sigma}^{(2)}=z.
\end{eqnarray}
Thus, we obtain a non singular symplectic matrix $\Tilde{f}_{ij}^{(2)}$ as
\begin{equation}
    \Tilde{f}_{ij}^{(2)}=\begin{pmatrix}
        0&-1&0&0&0&0&0&0\\
1&0&0&0&0&0&0&0\\
0&0&0&-1&0&0&0&0\\
0&0&1&0&0&0&g&0\\
0&0&0&0&0&-1&0&1\\
0&0&0&0&1&0&1&0\\
0&0&0&-g&0&-1&0&0\\
0&0&0&0&-1&0&0&0    
\end{pmatrix},
\end{equation}
and its inverse $\big(\Tilde{f}_{ij}^{(2)}\big)^{-1}$ is
\begin{equation}
    \big(\Tilde{f}_{ij}^{(2)}\big)^{-1}=\begin{pmatrix}
        0&1&0&0&0&0&0&0\\
-1&0&0&0&0&0&0&0\\
0&0&0&1&0&-g&0&-g\\
0&0&-1&0&0&0&0&0\\
0&0&0&0&0&0&0&-1\\
0&0&g&0&0&0&-1&0\\
0&0&0&0&0&1&0&1\\
0&0&g&0&1&0&-1&0    
\end{pmatrix}.
\end{equation}
According to Faddeev-Jackiw formalism, we can procure all the basic brackets from the components of the inverse symplectic matrix $\big(\Tilde{f}_{ij}^{(2)}\big)^{-1}$ as
\begin{eqnarray}\label{72}
    &&\big\{r,P_{r}\big\}=1=-\big\{P_{r},r\big\},\quad
     \big\{\theta,P_{\theta}\big\}=1=-\big\{P_{\theta},\theta\big\},\quad
     \big\{P_{z},\theta\big\}=g=-\big\{\theta,P_{z}\big\},\nonumber\\ 
       &&\big\{\sigma,\theta\big\}=g=-\big\{\theta,\sigma\big\},\quad
     \big\{\sigma,z\big\}=1=-\big\{z,\sigma\big\},\quad
      \big\{\rho,P_{z}\big\}=1=-\big\{P_{z},\rho\big\},\nonumber\\
      && \big\{\rho,\sigma\big\}=1=-\big\{\sigma,\rho\big\}.
    \end{eqnarray}
The brackets among the basic variables of the theory are exactly same as the Dirac brackets obtained via Dirac's formalism (cf. Appendix 1 for details).

\subsection{Gauge and (anti-)BRST Symmetries}
We have seen that, the first-iterated symplectic matrix is singular which indicates that underlying theory is a gauge theory. The zero-mode of the corresponding matrix is obtained as
\begin{equation}
    \big(\Tilde{\nu}^{(1)}\big)^{T}=\begin{pmatrix}
        0&0&-g&0&-1&0&1
    \end{pmatrix}.
\end{equation}
This zero-mode acts as the generator of the gauge transformations in the following manner:
\begin{equation}
    \delta\Tilde{\chi}_{k}^{1}=\Tilde{\nu}_{k}^{1}\lambda(t),
    \end{equation}
where $\Tilde{\chi}_{k}^{1}$ is set of symplectic variables and $\lambda(t)$ is the gauge parameter. Therefore, we can obtain the gauge transformations of each variable in the theory as
\begin{equation}
\begin{split}
    &\Tilde{\delta} r=0,\qquad \Tilde{\delta} P_{r}=0,\qquad
    \Tilde{\delta} \theta=-g\lambda(t),\\ &\Tilde{\delta} P_{\theta}=0,\qquad
    \Tilde{\delta} z=-\lambda(t),\qquad \Tilde{\delta} P_{z}=0, \qquad \Tilde{\delta}\zeta=-\dot{\lambda}(t).
    \end{split}
\end{equation}
The first-order Lagrangian $\Tilde{L}_{f}^{(0)}$ is invariant under the above set of gauge transformations.
Now, we look into the BRST symmetries of the FLPR model in polar coordinates. To procure all the BRST transformations, we introduce anti-commuting variables $\cal{C}$, $\Bar{\cal{C}}$ and a commuting variable $b$. Now, we can write the BRST transformations as
\begin{equation}\label{BRST}
\begin{split}
    &\Tilde{\delta}_{b} r=0,\qquad \Tilde{\delta}_{b} P_{r}=0,\qquad
    \Tilde{\delta}_{b} \theta=-g{\cal{C}},\qquad\Tilde{\delta}_{b} P_{\theta}=0,\qquad\Tilde{\delta}_{b} z=-{\cal{C}},\\&
     \Tilde{\delta}_{b} P_{z}=0, \qquad \Tilde{\delta}_{b}\zeta=-\dot{\cal{C}},\qquad
   \Tilde{\delta}_{b}{\cal{C}}=0,\qquad \Tilde{\delta}_{b}{\cal\bar{C}}=-b,\qquad\Tilde{\delta}_{b} b=0.
    \end{split}
\end{equation}
The anti-BRST transformations are
\begin{equation}\label{80}
\begin{split}
    &\Tilde{\delta}_{ab} r=0,\qquad \Tilde{\delta}_{ab} P_{r}=0,\qquad
    \Tilde{\delta}_{ab} \theta=-g{\cal\bar{C}},\qquad\Tilde{\delta}_{ab} P_{\theta}=0,\qquad\Tilde{\delta}_{ab} z=-{\cal\bar{C}},\\&
     \Tilde{\delta}_{ab} P_{z}=0, \qquad \Tilde{\delta}_{ab}\zeta=-{\dot{\bar{\cal{C}}}},\qquad
   \Tilde{\delta}_{ab}{\cal\bar{C}}=0,\qquad \Tilde{\delta}_{ab}{\cal{C}}=b,\qquad\Tilde{\delta}_{ab} b=0.
    \end{split}
\end{equation}
In the BRST formulation, we are fixing the gauge by adding an (anti-)BRST invariant function to the first order Lagrangian. So, we have
\begin{eqnarray}\label{ab}
    \Tilde{L}_{b}&=&\Tilde{L}_{f}^{(0)}+\Tilde{\delta}_{b}\Big(\bar{\cal{C}}(\dot{\zeta}-z-\frac{1}{2}b)\Big)\nonumber\\&\equiv& \Tilde{L}_{f}^{(0)}+\Tilde{\delta}_{ab}\Big(-{\cal{C}}(\dot{\zeta}-z-\frac{1}{2}b)\Big)\nonumber\\&\equiv&P_{r}\dot{r}+P_{\theta}\dot{\theta}+P_{z}\dot{z}-\frac{1}{2}p_{r}^{2}-\frac{1}{2r^{2}}p_{\theta}^{2}-\frac{1}{2}P_{z}^{2}-\zeta(gP_{\theta}+P_{z})\nonumber\\&-&V(r)-b(\dot{\zeta}-z-\frac{1}{2}b)-{\dot{\bar{\cal{C}}}}\dot{\cal{C}}-\bar{\cal{C}}{\cal{C}},
\end{eqnarray}
where $b$ represents the Nakanishi-Lautrup auxiliary variable which linearizes the quadratic gauge fixing term $\frac{1}{2}(\dot{\zeta}-z)^{2}$ and $({\bar{\cal{C}}}){\cal{C}}$ denote the (anti-)ghost variables.
This Lagrangian remains quasi-invariant under the set of (anti-)BRST transformations given in \eqref{BRST} and \eqref{80} respectively. The nilpotent and absolutely anti-commuting (anti-) BRST charges are given as follows:
\begin{equation}
    \Tilde{Q}_{b}={\cal{C}}(-gP_{\theta}-P_{z})-\dot{\cal{C}}P_{\zeta}, \qquad \Tilde{Q}_{ab}=\bar{\cal{C}}(-gP_{\theta}-P_{z})-\dot{\bar{\cal{C}}}P_{\zeta}.
\end{equation}
The conservation of the (anti-)BRST charges can be demonstrated by use of the following Euler-Lagrange equations of motion corresponding to $\Tilde{L}_{b}$ in \eqref{ab}
\begin{equation}
    \begin{split}
        &\dot{P_{r}}-\frac{P_{\theta}^{2}}{r^{3}}+\frac{\partial V}{\partial r}=0, \qquad \dot{r}-P_{r}=0,\qquad \dot{P_{\theta}}=0,\qquad \dot{\theta}-\frac{P_{\theta}}{r^{2}}-\zeta g=0,\\& \dot{\zeta}-b-z=0,\qquad \dot{P_{z}}-b=0,\qquad \dot{z}-P_{z}-\zeta=0,\\& \dot{b}-gP_{\theta}-P_{z}=0,\qquad \ddot{\cal{C}}-{\cal{C}}=0,\qquad {\ddot{\bar{\cal{C}}}}-\bar{\cal{C}}=0.
    \end{split}
\end{equation}
Moreover, the (anti-)BRST charges $(\Tilde{Q}_{(a)b})$ annihilate the physical subspace as 
\begin{equation}
    \Tilde{Q}_{(a)b}| \psi \rangle=0.
\end{equation}
This condition signifies that the physical states of the system are annihilated by first class constraints $(-gP_{\theta}-P_{z})\approx0$ and $P_{\zeta}\approx0$ which is compatible with the Dirac formalism for quantization of constrained systems.

\section{Conclusions}\label{4}
We have accomplished the quantization of the FLPR model in Cartesian as well as polar coordinates \textit{\`{a} la} Faddeev-Jackiw formalism. The constraints in theory have been deduced by employing modified Faddeev-Jackiw formalism. After  incorporating all the constraints into the Lagrangian, in both the cases, we end up with a two-form singular symplectic matrix. This clearly indicated the presence of gauge symmetries in the system. Thus, in order to quantize the system, we have chosen a well defined admissible gauge fixing condition \cite{12,14}. This results into the formation of a non-singular symplectic two-form matrix. All the basic brackets have been procured from the inverse of this matrix. 

\par Further, we have identified the zero-mode of the first-iterated singular symplectic matrix as the generator of gauge transformations and deduce the gauge symmetries in the system. Moreover, we have established off-shell nilpotent and absolutely anti-commuting (anti-)BRST symmetries of the theory. We have also derived the conserved (anti-)BRST charges with the aid of Noether's theorem. In addition, using the physicality condition of dynamically stable subspace, we have shown that the physical states of the theory are annihilated by the first class constraints which is consistent with the Dirac formalism of quantization of constrained systems.

\section*{Acknowledgements:} ASN would like to thank Anjali S for productive discussion. SG would like to dedicate this paper in the memory of Prof. G. Rajasekaran who was an excellent teacher, a great leader and an inspiring mentor. We would also like to thank our esteemed Reviewer for fruitful and enlightening comments.

\section*{Appendix 1: Constraint Structure and Dirac Brackets}
Here we analyse the FLPR model within the framework of Dirac formalism \cite{1,2}. We start with the Lagrangian in Cartesian coordinates (cf. \eqref{Lc}). We identify, \textit{\`{a} la} Dirac formalism, $P_{\zeta}\approx0 $ as the primary constraint in the theory from the expression of canonical momenta (cf. \eqref{Cm}) corresponding to Lagrangian (cf. \eqref{Lc}),  i.e., 
\begin{equation}
    \phi_{1}\equiv P_{\zeta}\approx 0.
\end{equation}
Applying the consistency condition on primary constraint using canonical Hamiltonian (cf. \eqref{Hc}), we obtain a secondary constraint as:
\begin{equation}
    \phi_{2}\equiv g(yP_{x}-xP_{y})-P_{z}\approx 0.
\end{equation}
Consistency condition on the secondary constraint leads to an identity which implies that there are no further constraints in the system. In addition, these two constraints are first class in nature. Thus, in order to fix the gauge completely, we need to choose two gauge conditions $\zeta=0 \; (\equiv\phi_{3})$ and $z=0 \; (\equiv\phi_{4})$. It worthwhile to mention that, in Faddeev-Jackiw quantization method, only one constraint $\Omega^{(0)}\equiv\nu_{\lambda}\Big(g(xP_{y}-yP_{x})+P_{z}\Big)$ appears and hence we need one condition to fix the gauge completely. Moreover, from the expression of symplectic matrix, we inferred that our choice of gauge condition should depend on at least one of the symplectic variables. Thus, we choose $z=0$ as the gauge fixing condition (since $\zeta$ is not a symplectic variable and hence it is inadequate to fix the gauge completely) in Faddeev-Jackiw formalism.

Coming back to the Dirac formalism, we construct the non-singular $4\times 4$ matrix with elements $M_{ij}=\{\phi_{i},\phi_{j}\}$ as follows:
\begin{equation}
    M = \begin{pmatrix}
        0&0&-1&0\\
        0&0&0&1\\
        1&0&0&0\\
        0&-1&0&0
    \end{pmatrix}.
\end{equation}
The Dirac bracket for any two variable $A$ and $B$, in this case, is defined as \cite{2}
\begin{equation}
    \{A, B\}_{D}=\{A,B\}-\sum_{i,j=1}^{4}\{A, \phi_{i}\}\big(M^{-1}\big)_{ij}\{\phi_{j}, B\}.
\end{equation}
Thus, we have following non-vanishing Dirac brackets
\begin{equation}\label{dc}
\begin{split}
    &\{x,P_{x}\}_{D}=1, \qquad \{y,P_{y}\}_{D}=1, \qquad \{x,P_{z}\}_{D}=gy, \\ &\{P_{z},y\}_{D}=gx, \qquad \{P_{x},P_{z}\}_{D}=gP_{y}, \qquad \{P_{z},P_{y}\}_{D}=gP_{x}.
    \end{split}
\end{equation}

\par Similarly, in polar coordinates, we start with the Lagrangian \eqref{43}. The canonical momenta as defined in \eqref{cmp}, indicate that $P_{\zeta}\approx0 \; (\equiv\Tilde{\phi}_{1})$ is a primary constraint in the theory according to the Dirac formalism for the classification of constraints. We deduce the secondary constraint using canonical Hamiltonian in \eqref{45}, as follows:
\begin{equation}
   \Tilde{\phi}_{2}\equiv-gP_{\theta}-P_{z}.
\end{equation}
There are no further constraints present in the theory. The vanishing Poisson bracket between these constraints indicates that these are first class constraints. Therefore, to fix the gauge and quantize the theory, we choose two admissible gauges $\zeta=0 \; (\equiv\Tilde{\phi}_{3})$ and $z=0 \; (\equiv\Tilde{\phi_{4}})$. Consequently, we obtain following Dirac brackets
\begin{equation}\label{dp}
\begin{split}
    &\{r,P_{r}\}_{D}=1, \qquad \{\theta,P_{\theta}\}_{D}=1, \qquad \{P_{z},\theta\}_{D}=g.
    \end{split}
\end{equation}
remaining brackets are zero. Thus, from this analysis, we observe that the Dirac brackets among the dynamical variables of the theory coincide with the basic brackets procured via Faddeev-Jackiw formalism (cf. \eqref{33} and \eqref{72}).

\section*{Appendix 2: On the Lagrange Multipliers}
In Faddeev-Jackiw formalism, Lagrange multipliers are introduced in order to incorporate the constraints into the Lagrangian. The basic brackets involving Lagrange multipliers did not appear in Dirac formalism as they are strong relation for the constraints \cite{9}. Here we provide a new interpretation for the Lagrange multipliers by solving the symplectic equations of motion. In case of Cartesian coordinates, the equations of motion corresponding to the Lagrangian $L_{f}^{(2)}$ can be expressed as follows:
\begin{equation}
    f_{ij}^{(2)}{\dot{\chi}^{(2)j}}=\frac{\partial V^{(2)}(\chi)}{\partial\chi^{(2)i}}\implies
    \dot{\chi}^{(2)j}=(f_{ij}^{(2)})^{-1}\frac{\partial V^{(2)}(\chi)}{\partial\chi^{(2)i}},
\end{equation}
\begin{equation}\label{38}
    \begin{pmatrix}
        \dot{x}\\
        \dot{P_{x}}\\
        \dot{y}\\
        \dot{P_{y}}\\
        \dot{z}\\
        \dot{P_{z}}\\
        \dot{\beta}\\
        \dot{\alpha}
    \end{pmatrix}=\begin{pmatrix}
        0&1&0&0&0&gy&0&gy\\
-1&0&0&0&0&gP_{y}&0&gP_{y}\\
0&0&0&1&0&-gx&0&-gx\\
0&0&-1&0&0&-gP_{x}&0&-gP_{x}\\
0&0&0&0&0&0&0&-1\\
-gy&-gP_{y}&gx&gP_{x}&0&0&-1&0\\
0&0&0&0&0&1&0&1\\
-gy&-gP_{y}&gx&gP_{x}&1&0&-1&0
    \end{pmatrix}
    \begin{pmatrix}
        2U^{'}x\\
        P_{x}\\
        2U^{'}y\\
        P_{y}\\
        0\\
        P_{z}\\
        0\\
        0
    \end{pmatrix}.
\end{equation}
Therefore, we have
\begin{equation}\label{lmc}
\begin{split}
    &\dot{x}=P_{x}+gyP_{z},\qquad \dot{P_{x}}=-2U^{'}x+gP_{y}P_{z},\qquad \dot{y}=P_{y}-gxP_{z},\\ &\dot{P_{y}}=-2U^{'}y-gP_{x}P_{z},\qquad \dot{z}=0,\qquad \dot{P_{z}}=0,\qquad \dot{\beta}=P_{z},\qquad \dot{\alpha}=0.
    \end{split}
   \end{equation}
    Here, one of the Lagrange multipliers can be identified as the generalised momenta. Using the above relations \eqref{lmc} and constraint equation \eqref{cc}, we can prove that our quantized results are precisely consistent.
    \par Similarly, in polar coordinates, the symplectic equations of motion corresponding to the Lagrangian \eqref{65}, are given as
\begin{equation}
    {\dot{\Tilde{\chi}}^{(2)j}}=(\Tilde{f}_{ij}^{(2)})^{-1}\frac{\partial \Tilde{V}^{(2)}(\Tilde{\chi})}{\partial\Tilde{\chi}^{(2)i}},
\end{equation}
\begin{equation}
\begin{pmatrix}
        \dot{r}\\
        \dot{P_{r}}\\
        \dot{\theta}\\
        \dot{P_{\theta}}\\
        \dot{z}\\
        \dot{P_{z}}\\
        \dot{\rho}\\
        \dot{\sigma}
    \end{pmatrix}=\begin{pmatrix}
        0&1&0&0&0&0&0&0\\
-1&0&0&0&0&0&0&0\\
0&0&0&1&0&-g&0&-g\\
0&0&-1&0&0&0&0&0\\
0&0&0&0&0&0&0&-1\\
0&0&g&0&0&0&-1&0\\
0&0&0&0&0&1&0&1\\
0&0&g&0&1&0&-1&0    
\end{pmatrix}
    \begin{pmatrix}
        -\frac{P_{\theta}^{2}}{r^{3}}+\frac{\partial V}{\partial r}\\
        P_{r}\\
        0\\
        \frac{P_{\theta}}{r^{2}}\\
        0\\
        P_{z}\\
        0\\
        0
    \end{pmatrix}.
\end{equation}
From the above relations, we obtain
\begin{equation}\label{lmp}
\begin{split}
    &\dot{r}=P_{r},\qquad \dot{P_{r}}=\frac{P_{\theta}^{2}}{r^{3}}-\frac{\partial V}{\partial r},\qquad \dot{\theta}=\frac{P_{\theta}}{r^{2}}-gP_{z},\\&\dot{P_{\theta}}=\dot{z}=\dot{P_{z}}=\dot{\sigma}=0,\qquad \dot{\rho}=P_{z}.
    \end{split}
    \end{equation}
    In this case also, one of the lagrange multipliers $\rho$ can be recognised as the generalised momenta. Moreover, it is straightforward to show that using \eqref{lmp} and constraint relation \eqref{cp} that our quantized results are consistent.

\end{document}